\documentclass[a4paper,11pt]{scrartcl}
\pdfoutput=1
\usepackage{amsmath,amssymb}
\usepackage{bbm}
\usepackage{graphicx}
\usepackage{multirow}

\addtokomafont{disposition}{\rmfamily\boldmath}
\let\tmptitle\title\renewcommand{\title}[1]{\tmptitle{\LARGE #1}}
\let\tmpauthor\author\renewcommand{\author}[1]{\tmpauthor{\large #1}}
\let\tmpdate\date\renewcommand{\date}[1]{\tmpdate{\normalsize #1}}
\newcommand{\abstrct}[1]{\begin{abstract}\vspace{-2em}\small\noindent#1\end{abstract}}

\title{\Large
CP Violation in Supersymmetry with Effective Minimal Flavour Violation
}
\date{}
\author{
Riccardo Barbieri, Paolo Lodone and David M. Straub
\\ \normalsize\itshape
Scuola Normale Superiore and INFN, Piazza dei Cavalieri 7, 56126 Pisa, Italy
}

\begin{document}

\maketitle

\abstrct{%
We analyze CP violation in supersymmetry with Effective Minimal Flavour Violation, as recently proposed in \cite{Barbieri:2010ar}. Unlike the case of standard Minimal Flavour Violation, we show that all the phases allowed by the flavour symmetry can be sizable without violating existing Electric Dipole Moment constraints, thus solving the SUSY CP problem. The EDMs at one and two loops are precisely analyzed as well as their correlations with the expected CP asymmetries in $B$ physics.
}

\section{Introduction}

If weak-scale supersymmetry plays a role in the extension of the Standard Model, given the plethora of potential new sources of flavour and CP violation present in the general MSSM, one of the most pressing questions to answer is why we have not seen  conclusive evidence for any of them so far. This is the essence of the SUSY flavour and CP problems.

As long as we are ignorant about the mechanism of supersymmetry and flavour symmetry breakings, a phenomenologically successful assumption that allows to address these issues is the principle of Minimal Flavour Violation (MFV) \cite{Hall:1990ac, Buras:2000dm,D'Ambrosio:2002ex}. The MFV principle assumes that all flavour violations are governed by the Yukawa couplings with suitable symmetry properties and thus evades the SUSY flavour problem. However, it does not by itself provide a solution to the SUSY CP problem \cite{Dugan:1984qf, Colangelo:2008qp,Mercolli:2009ns,Paradisi:2009ey,Ellis:2007kb}: flavour blind phases, such as the phase of the $\mu$ term, the gaugino masses and the trilinear couplings in the MSSM, are not forbidden and, unless strongly suppressed,  would violate bounds set by the non-observation of the electric dipole moments (EDMs) of the electron or neutron.

A different way to address the SUSY flavour problem,
that has received renewed interest recently \cite{Giudice:2008uk,Barbieri:2010ar}, is to assume a strong generational hierarchy in the squark mass spectrum \cite{Dine:1993np, Pouliot:1993zm,Pomarol:1995xc,Barbieri:1995uv,Cohen:1996vb,Dimopoulos:1995mi}.
This possibility is phenomenologically appealing since it might help evading the strong bounds from $K$ physics on
the flavour structure of the first two generations by making the corresponding squarks heavy. This without worsening the gauge hierarchy problem, given that the squarks of the third generation, which couple most strongly to the Higgs system, remain close to the electroweak scale.
Interestingly, decoupling the first generation squark masses also weakens the bounds on CP violating phases, since
the one-loop contributions to the experimentally accessible EDMs involve the superpartners of first generation fermions.
Thus, a hierarchy in the squark spectrum could in principle ameliorate both the flavour and the CP problems.	

Still, for a  generic flavour structure of the MSSM soft terms, a reasonably hierarchical squark spectrum is far from enough to solve the flavour problem, so extra symmetries are required \cite{Giudice:2008uk}.
An interesting possibility has been put forward in \cite{Barbieri:2010ar}, based on two assumptions. First, only those sfermions interacting with the Higgs system via the top Yukawa, i.e. the top squarks and the left-handed bottom squark are light. Second, in the limit of vanishing down-quark Yukawa couplings, $Y_d=0$, there is no flavour transition in the quark sector, i.e. the squark mass matrices in the up-type sector are aligned with the up-type Yukawa matrix. Such assumptions are robust in the sense that they are consistent with a suitable symmetry pattern, as specified below.
Since such symmetry does  not imply a degeneracy of the first and second generation squarks,
the dominant constraint arises from $K^0$-$\bar K^0$ mixing. Once this constraint is fulfilled by a sufficient hierarchy between the third and the first two generations,  it leads to Effective Minimal Flavour Violation (EMFV) in all other FCNC processes.

The aim of this work is to extend the discussion of this framework to allow all the CP violating phases not forbidden by the flavour symmetry, which have been neglected in \cite{Barbieri:2010ar} for simplicity.
As said, with hierarchical squark masses, the usual argument that these phases need to be tiny to meet the EDM bounds does not apply.
With the one-loop contributions to EDMs suppressed by first-generation squark masses, the most interesting signatures of EMFV are expected to arise in EDMs, from two loop effects, as well as in  $b\to s$ transitions,  sensitive to the exchange of the third generation of squarks only. This is in particular true for moderate values of $\tan{\beta}$, to which we stick throughout this work, since we view the down-type Yukawa couplings as a perturbation relative to the up-type ones.

Our strategy is as follows. After briefly describing the setup of EMFV in section \ref{sec:setup}, we discuss the EDM constraints in section~\ref{sec:edm}. In section~\ref{sec:bphys}, we consider the possible effects that can arise in $B$ physics observables. Section~\ref{sec:num} is devoted to the numerical analysis, confronting the EDM constraints with the signatures in $B$ physics. Section~\ref{sec:concl} contains our conclusions.

\section{Effective Minimal Flavour Violation}
\label{sec:setup}

In the absence of down-type Yukawa couplings, we assume   the following flavour symmetry in the quark sector  \cite{Barbieri:2010ar}
\begin{equation}
U(1)_{\tilde{B}_1}\times U(1)_{\tilde{B}_2}\times U(1)_{\tilde{B}_3}\times U(3)_{d_R},
\label{U3}
\end{equation}
where $\tilde{B}_i$ acts as baryon number but only on the supermultiplets $\hat{Q}_i$ and $\hat{u}_{R_i}$ of the i-th generation, respectively the left-handed doublets and the charge 2/3 right-handed singlets, whereas $U(3)_{d_R}$ acts on the three right-handed supermultiplets of charge $-1/3$.
Still with $Y_d=0$, in the physical basis for the up-type quarks and with every interaction flavour diagonal, this requires the following pattern in the squark mass matrices,
\begin{align}
m_Q^2 &=\text{diag}\,( m_{\tilde q_1}^2, m_{\tilde q_2}^2, m_{\tilde q_3}^2) ~,
\label{eq:mQ2}\\
m_U^2 &=\text{diag}\,( m_{\tilde u_1}^2, m_{\tilde u_2}^2, m_{\tilde u_3}^2) ~,
\\
m_D^2 &=  m_{\tilde d}^2 \times\mathbbm{1} ~.
\end{align}
Here, only $m_{\tilde q_3}^2$ and $m_{\tilde u_3}^2$ are assumed to be light, while the other mass squared parameters are heavy. We also assume all the slepton masses to be heavy. 

Once $Y_d$ is switched on, the flavour symmetry is broken down to baryon number and the squark mass matrices $m_Q^2$ and $m_D^2$ receive corrections quadratic in $Y_d$, since, in analogy with standard MFV, we promote $Y_d$ to a non-dynamical spurion field transforming under (\ref{U3}) in such a way that the down-Yukawa couplings are formally invariant.
As mentioned above, we 
assume $\tan\beta$ to be small to moderate, in the range $\tan\beta\lesssim 5$,
throughout this work.
To see the effects of the corrections induced by $Y_d$, one must view $Y_d$ as the sum of 3 matrices $Y_d^i$, each with only the $i$-th row different from zero,
\begin{equation}
Y_d^i=\mathbbm{1}^i Y_d \,,\qquad (\mathbbm{1}^i)_{\alpha\beta}=\delta_{i\alpha}\delta_{i\beta}\,.
\end{equation}
The $Y_d^i$ transform as a triplet under $U(3)_{d_R}$ and are charged under $\tilde B_i$.\footnote{We thank Marco Nardecchia for pointing this out to us.}
It follows that
\begin{align}
\Delta m_Q^2 &= \sum_i m_{\tilde q_i}^2 Y_d^i(Y_d^i)^\dagger \,, \\
\Delta m_D^2 &= \sum_i m_{\tilde d_i}^2 Y_d^i(Y_d^i)^\dagger \,,
\end{align}
where $ m_{\tilde q_i}^2$, $m_{\tilde d_i}^2$ are real squared masses. Therefore, $m_Q^2$ remains diagonal, whereas, setting $Y_d=V \hat Y_d$ by $U(3)_{d_R}$ invariance, with $V$ the CKM matrix and $\hat Y_d$ diagonal, the correction to $m_D^2$ becomes
\begin{equation}
\Delta m_D^2 = \sum_i m_{\tilde d_i}^2 \hat Y_d V^\dagger \mathbbm{1}^i V \hat Y_d\,,
\end{equation}
which is negligibly small for $m_{\tilde d_i}^2=O(m_{\tilde d}^2)$.

In the physical basis for all matter fields, quarks and squarks, all the flavour changing interaction terms are therefore
\begin{eqnarray}
\mathcal{L}_{FC} &=& \frac{g}{\sqrt{2}}\, (\overline{u_L} \gamma^\mu V\, d_L) \, W^+_\mu - g\, \tilde{u}^*_L V \,\overline{\tilde{W}^- }\, d_L \,+ \frac{g}{\sqrt{2}}\, \tilde{d}_L^* V \, \overline{\tilde{W}^3}\, d_L \nonumber \\
&& -\sqrt{2} \frac{g'}{6} \tilde{d}^*_L\, V \,\overline{\tilde{B}}\, d_L -\sqrt{2}\, g_3 \, \tilde{d}_L^* \, \lambda^b\, V\, \overline{\tilde{g}^b }\,d_L \nonumber \\
&& + \tilde{{u}}_R^* \, \hat Y_u \, V \, \overline{\tilde{H}_{u}^- }\, d_L+\overline{{u}_R} \, \hat Y_u\, V \,d_L \, H_u^+ + h.c.,\nonumber \\
\label{LFC}
\end{eqnarray}
where $\hat Y_u$ is the diagonal Yukawa coupling matrix and terms proportional to $Y_d$ have been neglected.

Concerning the trilinear couplings, the flavour symmetry (\ref{U3}) forbids $A$ terms for the down-type squarks, but only requires the up-type trilinears to be diagonal in the basis where the up-type Yukawas are diagonal, with no restriction on their size or phases. This is in contrast to the MFV case, where the first two generation $A$ terms are proportional to the first two generation Yukawas. However, this will not play an important role in the following, since the heaviness of the first two generation squarks implies that left-right mixing is always a small effect, except for the stop.
The $A$ terms of the down-type squarks, on the other hand, have the MFV form
\begin{equation}
A_D = a_D Y_d + O(Y_d^3)~,
\label{eq:Ad}
\end{equation}
with $a_D$ in general complex.
Since all down-type squarks with the exception of the left-handed sbottom are heavy and since $\tan\beta$ is small, down-type trilinears will also be negligible for phenomenology. We can therefore restrict our discussion of trilinear couplings to the stop trilinear $A_t$ in the following\footnote{%
We do not factor out the top Yukawa from $A_t$ and choose a convention where the left-right mixing entry of the stop mass matrix is given by $\frac{v_u}{\sqrt{2}}(A_t-\mu^* y_t \cot\beta)$.
}.

As shown in \cite{Barbieri:2010ar},
in the absence of CP violating phases beyond the CKM phase,
the most important constraint on this setup arises from $K^0$-$\bar K^0$ mixing,
since
the non-degeneracy of the first two generation squark masses
leads to sizable gluino contributions to the $\Delta S=2$ effective Hamiltonian. However the fact that only the standard CKM matrix determines any flavour transition gives relatively mild constraints on the masses of the first two generations of squarks, which have to be above a few TeV.
Due to the  theory uncertainty on the SM value of the $K^0$-$\bar K^0$ mass difference $\Delta m_K$, it was found that the constraint coming from $|\epsilon_K|$ is typically stronger than the one coming from $\Delta m_K$.
Interestingly, the Standard Model prediction for $|\epsilon_K|$  has recently decreased due to improvements in the lattice QCD calculation of the bag parameter $\hat B_K$, leading to a lower central value, as well as previously neglected contributions beyond the lowest order in the operator product expansion \cite{Buras:2010pza}. As a result, the current prediction reads \cite{Brod:2010mj}
\begin{equation}
|\epsilon_K|^\text{SM} = (1.90 \pm 0.26) \times10^{-3} ~,
\label{e_th}
\end{equation}
which is $1.3\sigma$ below the experimental value
\begin{equation}
|\epsilon_K|^\text{exp} = (2.228 \pm 0.011)\times10^{-3} ~.
\label{e_exp}
\end{equation}
One can speculate that  the positive contributions to $|\epsilon_K|$ that are unavoidably generated in this framework can  explain the mismatch between the central values of (\ref{e_th}) and (\ref{e_exp}) for suitable values of the first and second generations of squarks masses\footnote{
 Since the SM prediction in (\ref{e_th}) sensitively depends on the relevant CKM parameters, alternative NP explanations would be a non-standard contribution to $B_d$ mixing or to the ratio $\Delta M_d/\Delta M_s$ (see e.g. \cite{Altmannshofer:2009ne}), which are however SM-like in this framework.
}.

Let us now consider which physical CP-violating phases are present in this framework. For simplicity we assume for the time being that the gaugino masses are universal, or at least have a common phase, at some scale (the implications of relaxing this assumption will be commented on in section~\ref{sec:num}). Then, by appropriate field redefinitions, one can choose the soft SUSY breaking $b$ term and the gaugino masses to be real. The remaining irreducible phases then reside in the $\mu$ term,  $\phi_\mu$, in the $a_D$ parameter of (\ref{eq:Ad}) as well as in the three up-squark trilinear couplings $A_{u,c,t}$. As discussed above, the only phenomenologically relevant phases will be the ones of $\mu$ and $A_t$, since the others are always accompanied by a heavy sfermion mass suppression.

\section{Electric Dipole Moments}
\label{sec:edm}

The non-observation of electric dipole moments of fundamental fermions is one of the strongest constraints on CP violation in the MSSM.
Experimentally, the most constraining EDMs are currently the ones of the Thallium and Mercury atoms and of the neutron.
The Thallium EDM is dominated by the electron EDM and is approximately given by
\begin{equation}
d_\text{Tl} = -585  \, d_e \,,
\end{equation}
leading to the experimental 90\% C.L. upper bound \cite{Regan:2002ta}
\begin{equation}
|d_e| < 1.6 \times 10^{-27} ~e\,\text{cm\,.}
\label{eq:deexp}
\end{equation}
The neutron EDM, on the other hand, receives contributions from the electric and chromoelectric dipole moments (CEDMs) of the up and down quarks. 
For the case of the neutron EDM one can use QCD sum rules \cite{Pospelov:2000bw} to get:
\begin{equation} \label{eq:QCDsumrules}
d_n = (1 \pm 0.5) \left[ \frac{\left< \overline{q}q \right>}{(225\mbox{ MeV})^3} \right] \left( 1.4( d_d - \frac{1}{4} d_u) + 1.1 e ( \tilde{d}^C_d + \frac{1}{2} \tilde{d}^C_u)  \right) \, ,
\end{equation}
where $\tilde{d}^C$ are the CEDMs, and chiral theory to write:
\begin{equation} \label{eq:chiraltheory}
\left< \overline{q}q \right> = \frac{f_\pi^2 m_{\pi^0}^2}{m_u + m_d} \, .
\end{equation}

The current experimental upper bound at the 90\% confidence level is \cite{Baker:2006ts}
\begin{equation}
|d_n| < 2.9 \times 10^{-26} ~e\,\text{cm\,.}
\label{nEDM_exp}
\end{equation}
The Mercury EDM is sensitive to the electron EDM and the quark CEDMs. In our framework it turns out that, also in view of considerable hadronic uncertainties \cite{Ellis:2011hp}, it is not competitive with the other two constraints, so we focus on $d_n$ and $d_e$ from now on.

\subsection{One loop EDMs}

\begin{figure}[tp]
 \centering
 \includegraphics[width=0.4\textwidth]{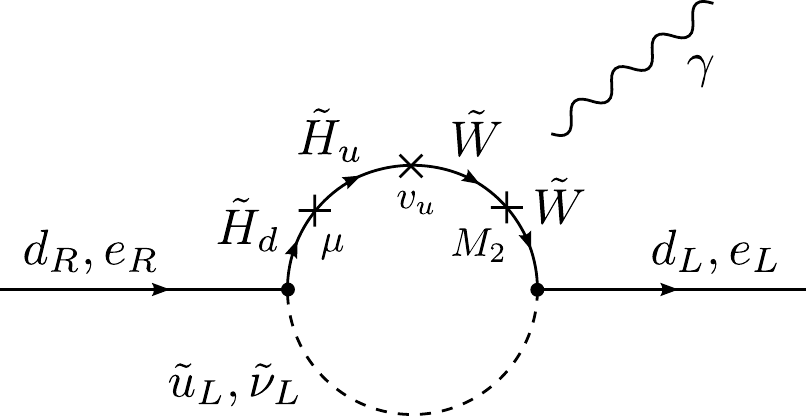}
 \caption{Dominant one-loop contribution to the down-quark and electron EDMs. The photon can be attached anywhere along the chargino line.}
 \label{fig:edmd}
\end{figure}

The one-loop contributions to the quark (C)EDMs are always suppressed by the heavy masses since the contributions involving only the third generation are suppressed by a factor of $|V_{td}|^2/|V_{ud}|^2\approx8\times10^{-5}$, which is significantly smaller than the generational suppression $m_l^2/m_h^2$ for the range of parameters we consider\footnote{Here and in the following, $m_h$ will denote the scale of the heavy sfermions as described in section~\ref{sec:setup}, while $m_l$ is the scale of the light sfermions, Higgses and fermionic sparticles.}. Indeed, the only diagram which is suppressed only by two powers of the ratio $m_l/m_h$
is the Higgsino-Wino contribution to the electron and the down quark EDM shown in figure~\ref{fig:edmd}, with the photon  attached to the Higgsino-Wino line even in the case of the $d$-quark. An analogous contribution exists for the up quark, but there the factor $\tan\beta$ has to be replaced by $\cot\beta$. In addition, the up-quark EDM enters the neutron EDM with a factor of $1/4$ with respect to the down quark as shown in (\ref{eq:QCDsumrules}). Therefore, $d_e$ and $d_d$ are more constraining.

Neglecting terms of order higher than 2 in the ratio  $m_l/m_h$, the contribution to $d_e$ and $d_d$ from the diagram in figure~\ref{fig:edmd} is
\begin{align}
\frac{d_{e,d}^{1,\tilde H \tilde W}}{e} &= 
\frac{\alpha}{4\pi\sin^2{\theta_W}} \;
\frac{m_{e,d}\,\tan\beta}{m_{\tilde \nu_L,\tilde u_L}^2}
\sin{(\phi_\mu)}
f_1\!\left(\frac{|M_2|}{|\mu|}\right)
\,,
\label{EDMs}
\end{align}
where 
\begin{equation}
f_1(x) = \frac{2x \ln x}{x^2-1} \,.
\end{equation}
For the case of the electron EDM, with $|\mu| = |M_2|$ the experimental limit  (\ref{eq:deexp}) is satisfied for
\begin{equation}
m_{\tilde{\nu}} >  \mathbf{4.0 \mbox{ \bf TeV }} \times (\sin \phi_\mu \tan \beta)^{\frac{1}{2}}.
\end{equation}

For the case of the neutron EDM, care must be taken to  account properly for the QCD running effects from the scale of the heavy squark exchanged in fig.~\ref{fig:edmd}   down to 1 GeV where the various quark terms in (\ref{eq:QCDsumrules}) are understood. To this end two considerations hold:
\begin{itemize}
\item At the high scale $m_h$, integrating out the heavy $\tilde{u}$ one does not generate an EDM since the quarks can still be considered massless, but one generates the operators (including the coefficients at $m_h$):
\begin{equation}
\Delta \mathcal{L} 
= \frac{g y_d|_{m_h}}{m^2_{ \tilde{u} }} \left[ \frac{1}{2}(\overline{d}_L d_R)(\overline{\tilde{H}}_{d\, L} \tilde{W}) + \frac{1}{8} (\overline{d}_L \sigma^{\mu\nu} d_R)(\overline{\tilde{H}}_{d\, L} \sigma_{\mu\nu}\tilde{W}) \right].
\label{eff_ops}
\end{equation}
Below $m_h$ these operators do not mix and the second operator in the r.h.s. of (\ref{eff_ops}) (the only one that contributes  at the weak scale where $v$ appears and one integrates out $\tilde{W}$ and $\tilde{H}_d$ to generate the EDM) runs with the same anomalous dimension of the EDM operator itself, $\gamma_\text{EDM} = 8/3 (\alpha_S/4\pi)$\cite{Degrassi:2005zd}.
\item  From (\ref{eq:QCDsumrules}) and (\ref{eq:chiraltheory}), the best way to estimate the neutron EDM is to consider the running of $d_q/m_q$ with anomalous dimension $\gamma = 32/3 (\alpha_S/4\pi)$ and use $m_u / m_d = 0.553 \pm 0.043$.
\end{itemize}
To include QCD running effects, therefore, the proper factor that multiplies $d_{d}^{1,\tilde H \tilde W}/m_d$ in (\ref{EDMs}) before its inclusion in (\ref{eq:QCDsumrules}) is
\begin{equation} \label{eq:QCDcorrectionFactor}
\eta_\text{QCD} = \left( \frac{\alpha_s(m_{\tilde{u}})}{\alpha_s(m_l)} \right)^{\frac{32/3}{2(9/2)}} \left( \frac{\alpha_s(m_l)}{\alpha_s(m_t)} \right)^{\frac{32/3}{2(7)}}  \left( \frac{\alpha_s(m_t)}{\alpha_s(m_b)} \right)^{\frac{32/3}{2(23/3)}} \left( \frac{\alpha_s(m_b)}{\alpha_s(1\mbox{ GeV})} \right)^{\frac{32/3}{2(25/3)}} \, ,
\end{equation}
where $m_l$ is the common mass of all the ``light'' s-particles and the different thresholds are taken into account in the $\beta$-function for $\alpha_s$. From (\ref{nEDM_exp}) and  the central value of (\ref{eq:QCDsumrules}) one gets the bound
\begin{equation}
m_{\tilde{u}} >  \mathbf{2.7 \mbox{ \bf TeV }} \times (\sin \phi_\mu \tan \beta)^{\frac{1}{2}}
\end{equation}
or, more conservatively, $m_{\tilde{u}} > 1.9~\text{TeV}~ (\sin \phi_\mu \tan \beta)^{\frac{1}{2}}$, if one uses the weaker constraint.

As anticipated, for the moderate values of $\tan\beta$ we consider, these constraints allow an arbitrarily large phase of the $\mu$ term for first generation sfermion masses which are perfectly natural in the framework described in \cite{Barbieri:2010pd}. This is at variance with the standard MFV case, where several one loop diagrams contribute to the EDMs 
and, taking all the s-particle masses at $\tilde{m}$ and a universal trilinear coupling $A_0$, the following bounds have to be satisfied:
\begin{itemize}
\item From the electron EDM
\begin{equation}
\mathbf{ \sin \phi_\mu \tan \beta < 7 \times 10^{-2}} \, \left( \frac{\tilde{m}}{500 \mbox{ GeV}} \right)^2 .
\label{eq:phimubd}
\end{equation}
\item From the neutron EDM (central value)
\begin{equation}
\sin \phi_{A_0} < \mathbf{  2 \times 10^{-1}} \, \left( \frac{\tilde{m}}{500 \mbox{ GeV}} \right)^2 \left( \frac{\tilde{m}}{|A_0|} \right) .
\label{eq:phiAbd}
\end{equation}
\end{itemize}

In fact, the bounds on $\sin \phi_\mu$ and $\sin \phi_{A_0}$ are even up to an order of 
magnitude stronger in big parts of the MFV parameter space than the bounds 
quoted  in eqs.~(\ref{eq:phimubd}) and (\ref{eq:phiAbd}), which are affected by accidental cancellations 
occuring in the case of degenerate s-particles at $\tilde{m}$.

\subsection{Two loop EDMs}

\begin{figure}[tp]
 \centering
 \includegraphics[width=\textwidth]{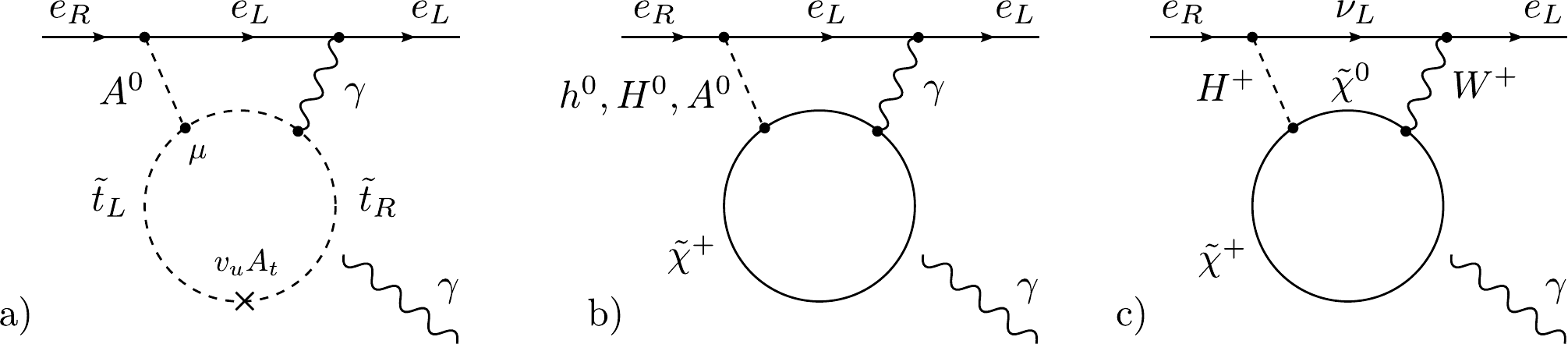}
 \caption{Two-loop Barr-Zee type diagrams contributing to the electron EDM. The photon can be attached anywhere along the loop.}
 \label{fig:edm-2loop}
\end{figure}

Due to the strong suppression of the one loop EDMs,
 the  two loop contributions come into play. Indeed, at two loop level there are Barr-Zee type diagrams not involving any of the first two generation squarks \cite{Chang:1998uc,Giudice:2005rz,Li:2008kz,Ellis:2008zy,Abel:2001vy}, such that the additional loop suppression can be compensated by the absence of the mass suppression. Some of these contributions are shown in figure~\ref{fig:edm-2loop} for the case of the electron EDM. As a matter of fact all the diagrams missing from  figure~\ref{fig:edm-2loop}
 are suppressed by a relative factor $1/\tan^2{\beta}$.\footnote{%
These very same diagrams are the ones that contribute in split supersymmetry where only one Higgs doublet survives in the spectrum at the Fermi scale \cite{Giudice:2005rz}.}.
 Analogous diagrams exist for the up and down quarks. However, the current experimental situation makes the Barr-Zee contribution to $d_e$ by far the most constraining one. 
 
\begin{figure}[tbp]
\begin{center}
 \raisebox{1cm}{a)}\includegraphics[width=0.4\textwidth]{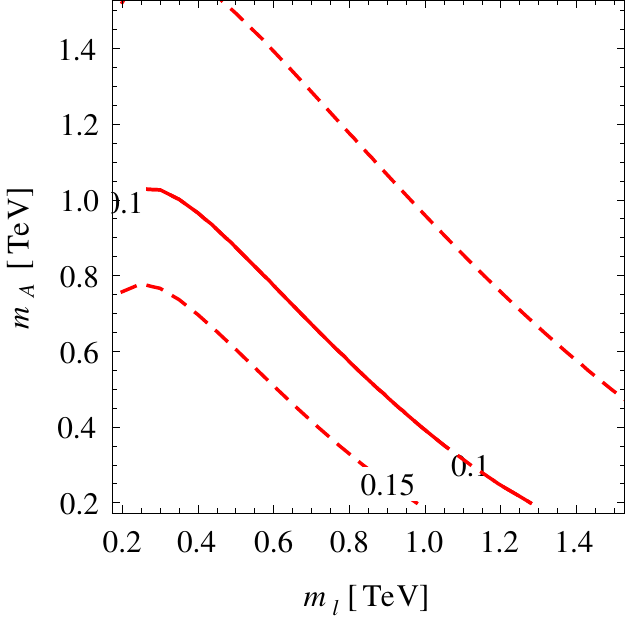}
 \hspace{1cm}
 \raisebox{1cm}{b)}\includegraphics[width=0.4\textwidth]{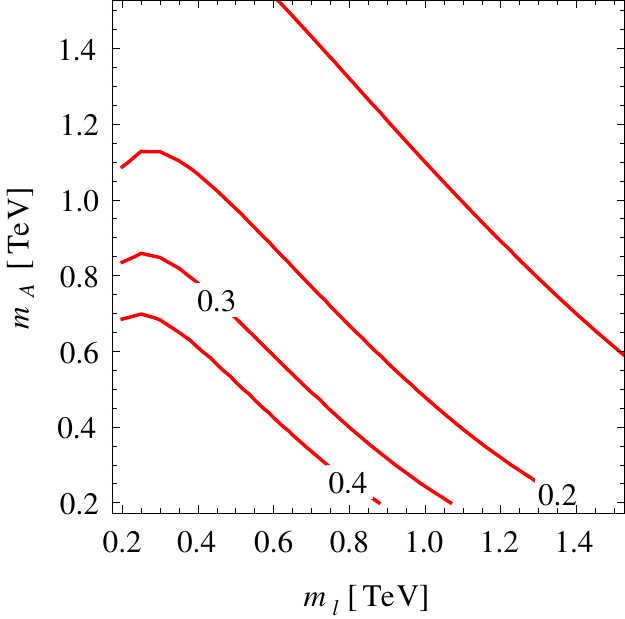}\\
 \raisebox{1cm}{c)}\includegraphics[width=0.4\textwidth]{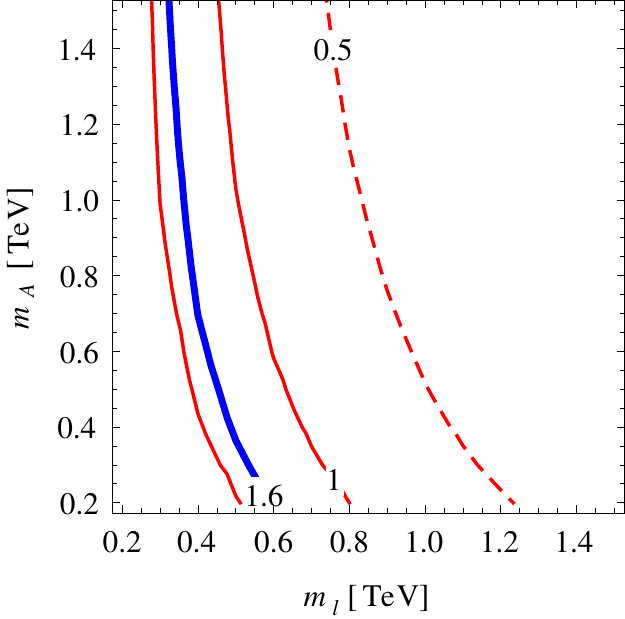}
 \hspace{1cm}
 \raisebox{1cm}{d)}\includegraphics[width=0.4\textwidth]{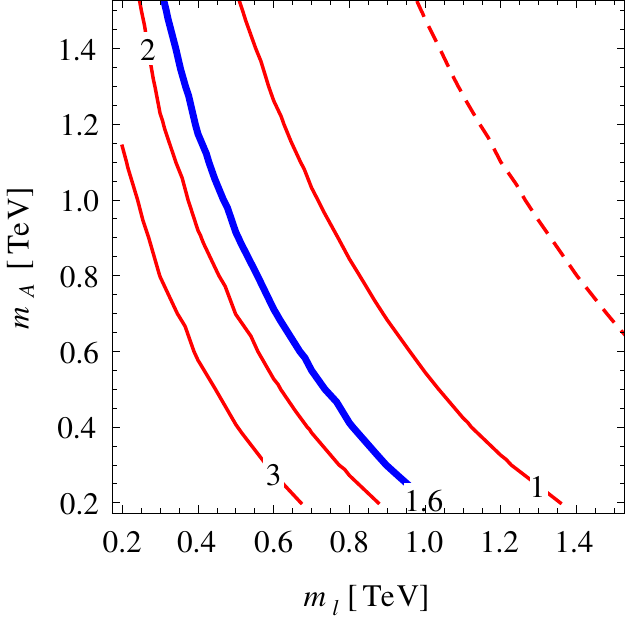}
\end{center}
\caption{Prediction for the electron EDM in units of $10^{-27}\;e\,$cm
in terms of the common stop and chargino mass $m_l$ and the common mass $m_A$ of $H^0$, $A^0$, $H^\pm$
in a scenario with $\sin{(\phi_{A_t}})=1$ (diagrams a and b) or $\sin{(\phi_\mu})=1$ (diagrams c and d) for $\tan\beta=2$ (a and c) and  $\tan\beta=5$ (b and d). The thick blue line in the lower plots corresponds to the 90\% C.L. experimental upper bound, with the area left of it excluded.
}
 \label{fig:edm-contours}
\end{figure}

Assuming the stop and chargino masses to be degenerate at $m_l$, and taking $H^0$, $A^0$, $H^\pm$ all at a common mass $m_A$, these two loop contributions to $d_e$ are shown in figures~\ref{fig:edm-contours}a)--d) for maximal, independent values of $\sin{(\phi_\mu})$ and $\sin{(\phi_{A_t}})$ and $\tan{\beta} = 2,5$.
Such contributions are irreducible in the sense that they do not decouple with the first two generation squark masses. However, it is interesting that, for $O(1)$ phases and natural values of all the relevant parameters, the prediction for $d_e$ is in the ballpark of the current experimental bound, eq. (\ref{eq:deexp}). We thus conclude that large flavour blind phases are allowed in EMFV, but predict an electron EDM in the reach of future experiments.

\section{CP asymmetries in $B$ physics}
\label{sec:bphys}

In addition to generating EDMs, the flavour blind phases also generate CP asymmetries in $B$ physics. Remarkably, these contributions are unsuppressed by the heavy generations in $B$ physics, which involves the third generation. 

The most relevant effects in EMFV are generated through contributions to the magnetic and chromomagnetic penguin operators in the $b\to s$ effective Hamiltonian
\begin{equation}
\mathcal H_\text{eff} = -\frac{4 G_F}{\sqrt{2}}V_{tb}V_{ts}^* \left(C_7 \mathcal{O}_7+C_8 \mathcal{O}_8\right)\,,
\end{equation}
\begin{align}
{\mathcal{O}}_{7} &= \frac{e}{16\pi^2} m_b
(\bar{s} \sigma_{\mu \nu} P_{R} b) F^{\mu \nu} ,&
{\mathcal{O}}_{8} &= \frac{g_3}{16\pi^2} m_b
(\bar{s} \sigma_{\mu \nu} T^a P_{R} b) G^{\mu \nu \, a} .&
\label{eq:DF1-1}
\end{align}

$\Delta B=2$ processes, on the other hand, are only weakly affected. In particular, a sizable phase in $B_s$ mixing, 
probed in the mixing-induced CP asymmetry in $B_s\to J/\psi\phi$ and in the like-sign dimuon charge asymmetry and currently favoured by the data \cite{Lenz:2010gu}, cannot be accommodated. As already mentioned, the only relevant effect in $\Delta F=2$ transitions can occur in $\epsilon_K$.

Before discussing the observables sensitive to the (chromo)magnetic operators,
let us discuss the
individual contributions to the Wilson coefficients $C_7$ and $C_8$
generated in the EMFV framework.

\subsection{Contributions to the magnetic and chromomagnetic operators}
\label{sec:C78}

\begin{figure}[tp]
 \centering
 \includegraphics[width=0.73\textwidth]{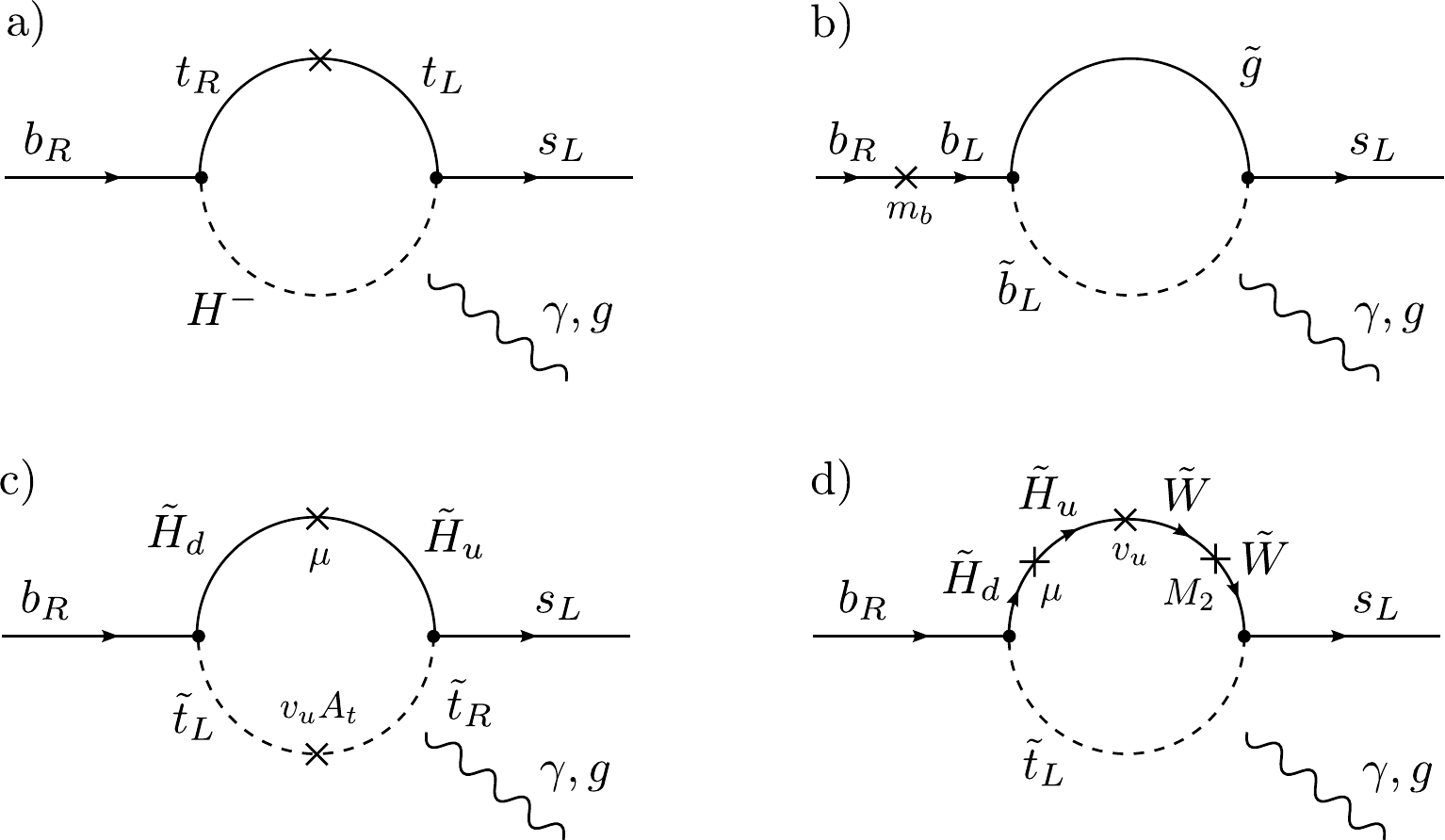}
 \caption{Main contributions to $C_7$ and $C_8$.}
 \label{fig:bsgamma}
\end{figure}

In the MSSM, the one-loop contributions to  $C_7$ and $C_8$ stem from charged Higgs/top, neutralino/down squark, chargino/up squark and gluino/down squark loops. In our framework, the neutralino contributions are subleading. The dominant effects are therefore generated by diagrams involving a charged Higgs, gluino or chargino and no sfermi\-ons besides the stops and left-handed sbottom.

The charged Higgs contribution shown in figure~\ref{fig:bsgamma}a) is given by
\begin{equation}
C_{7,8}^{H^\pm}= f_{7,8}\!\left(\frac{m_{H^\pm}^2}{m_{t}^2}\right)
\,,
\end{equation}
where $f_7(1)=-\frac{7}{36}$
and  $f_8(1)=-\frac{1}{6}$,
and is independent of $\tan\beta$.

The only gluino diagram not suppressed by a heavy mass is the one in figure~\ref{fig:bsgamma}b). Assuming the gluino and left-handed sbottom masses to be degenerate at $m_l$, it gives a contribution
\begin{equation}
\frac{4 G_F}{\sqrt{2}}\;C_{7,8}^{\tilde g}=
-
\frac{8}{3} \frac{g_s^2}{m_l^2}
\left\lbrace
\frac{1}{144},\frac{5}{144}
\right\rbrace
\,.
\end{equation}
The gluino contribution to $C_7$ is thus usually negligible with respect to the charged Higgs contribution due to the small loop function, while the contribution to $C_8$ can become relevant.

The chargino diagrams in figures~\ref{fig:bsgamma}c) and d) both involve a factor of $\tan\beta$ and become competitive with the charged Higgs contribution even for $\tan\beta$ as low as 5. Assuming the stop and chargino masses to be degenerate at $m_l$, the Higgsino diagram in figure~\ref{fig:bsgamma}c) can be approximately written as
\begin{equation}
\frac{4 G_F}{\sqrt{2}}\; C_{7,8}^{\tilde H}=
-
y_t\frac{\mu A_t}{m_l^4}
\tan\beta \;
\left\lbrace
\frac{5}{72},\frac{1}{24}
\right\rbrace
\,,
\end{equation}
while the diagram with Higgsino-Wino mixing in figure~\ref{fig:bsgamma}d) reads approximately
\begin{equation}
C_{7,8}^{\tilde H\tilde W}=
2m_W^2
\; \frac{\mu M_2}{m_l^4}
 \tan\beta \;
\left\lbrace
\frac{11}{72},\frac{1}{24}
\right\rbrace
\,.
\label{eq:C7HW}
\end{equation}

While the charged Higgs and gluino contributions are real\footnote{In the case of large $\tan\beta$, non-holomorphic corrections to the Yukawa couplings become relevant which can introduce phases in the charged Higgs contribution. Since we consider only low $\tan\beta$, we can neglect these corrections.}, both chargino diagrams contain irreducible CP violating phases.

We see that $C_7$ and $C_8$ can be significantly modified with respect to their SM values and they can acquire sizable phases, irrespective of the masses of the first two generation sfermions. The observables constraining these NP effects will be discussed in the next subsection.

\begin{table}[tp]
\renewcommand{\arraystretch}{1.4}
 \begin{center}
\begin{tabular}{l|lll}
Observable & SM prediction & Experiment & Future sensitivity\\
\hline
$\text{BR}(B\to X_s\gamma)$ & $(3.15\pm0.23)\times10^{-4}$ & $(3.52\pm0.25)\times10^{-4}$ & $\pm0.15\times10^{-4}$  \\
$A_\text{CP}(b\to s\gamma)$ & $\left(0.44^{+0.24}_{-0.14}\right) \%$ \cite{Hurth:2003dk}  & $(-1.2 \pm 2.8) \%$ & $\pm0.5\%$  \\
$S_{\phi K_S}$  & $0.68\pm0.04$ \cite{Buchalla:2005us,Beneke:2005pu} & $0.56^{+0.16}_{-0.18}$ & $\pm0.02$ \\
$S_{\eta' K_S}$ & $0.66\pm0.03$ \cite{Buchalla:2005us,Beneke:2005pu} & $0.59\pm0.07$  & $\pm0.01$ \\
$\langle A_7 \rangle$ & $(3.4\pm0.5)\times10^{-3}$ \cite{Altmannshofer:2008dz} & -- & ? \\
$\langle A_8 \rangle$ & $(-2.6\pm0.4)\times10^{-3}$ \cite{Altmannshofer:2008dz} & -- & ? \\
 \end{tabular}
 \end{center}
\renewcommand{\arraystretch}{1}
 \caption{SM predictions, current experimental world averages \cite{Barberio:2008fa} and experimental sensitivity at planned experiments \cite{Aushev:2010bq,O'Leary:2010af} for the $B$ physics observables. For the SM prediction of $A_\text{CP}(b\to s\gamma)$, note the comment on page~\pageref{acp}.}
 \label{tab:exp}
\end{table}

\subsection{Observables}

We now turn to the discussion of observables sensitive to NP effects in the Wilson coefficients of the magnetic and chromomagnetic operators. Apart from the branching ratio of $B\to X_s\gamma$, we focus on the CP asymmetries in $B\to X_s\gamma$ and $B\to K^*\mu^+\mu^-$ as well as the time-dependent CP asymmetries in $B\to\phi K_S$ and $B\to\eta' K_S$. Whereas $B\to K^*\mu^+\mu^-$ will be measured at the LHCb experiment, the other observables require the clean environment of an $e^+e^-$ machine and are going to be measured at the planned super flavour factories Belle~II and Super$B$. The current theoretical and experimental status and projected sensitivities are collected in table~\ref{tab:exp}. The theoretical uncertainties in this table can be somewhat optimistic, as explicitly commented below in the case of $A_\text{CP}(b\to s\gamma) $.

\subsubsection*{The $b\to s\gamma$ branching ratio and direct CP asymmetry}

The $B\to X_s\gamma$ branching ratio is a strong constraint on the overall NP contributions to $C_{7,8}$, in view of the good agreement between the SM prediction and the experimental measurement shown in table~\ref{tab:exp}.
In terms of the Wilson coefficients, the ratio of the branching ratio and its SM expectation reads
\begin{equation}
R_{b\to s\gamma} =
\frac{\text{BR}(B\to X_s\gamma)}{\text{BR}(B\to X_s\gamma)^\text{SM}}
=
\left| 1+\frac{C_7^\text{NP}}{C_7^\text{eff,SM}} \right|^2
\,,
\end{equation}
with the Wilson coefficients evaluated at the scale $m_b$.

The imaginary parts of $C_7$ and $C_8$ are probed by the direct CP asymmetry in $b\to s\gamma$,
\begin{equation}
A_\text{CP}(b\to s\gamma) =
\frac{\Gamma(\bar B\to X_s\gamma)-\Gamma(B\to X_{\bar s}\gamma)}{\Gamma(\bar B\to X_s\gamma)+\Gamma(B\to X_{\bar s}\gamma)}
\,.
\end{equation}
A simple expression for $A_\text{CP}$ can be obtained by multiplying it with $R_{b\to s\gamma}$.
Then, neglecting the small SM contribution and using the values given in \cite{Kagan:1998bh}, we find
\begin{equation}
A_\text{CP}(b\to s\gamma) \times R_{b\to s\gamma} =
-0.29 \,\text{Im}\left(C_7^\text{NP}\right)
+0.30 \,\text{Im}\left(C_8^\text{NP}\right)
-0.99 \,\text{Im}\left(C_7^\text{NP*}C_8^\text{NP}\right)
,
\end{equation}
where the Wilson coefficients are to be evaluated at the scale $m_b$.

As can be seen from table~\ref{tab:exp}, the experimental bound still leaves a large  room for new physics. The Belle-II collaboration aims to measure $A_\text{CP}$ to a precision of 0.5\%.

Recently,\label{acp} it has been pointed out that long distance effects that arise as corrections of order $\Lambda_\text{QCD}/m_b$ to the CP asymmetry might dominate the SM contribution and lead to sizable theory uncertainties \cite{Benzke:2010tq}. In particular, the authors of \cite{Benzke:2010tq} find $-0.5\%<A_\text{CP}^\text{SM}<2.6\%$. Due to the large uncertainty inherent in these contributions, we will not take them into account in our NP analysis, but keep in mind that significant theoretical progress will be required to interpret improved measurements of the CP asymmetry\footnote{%
We also note that the CP asymmetry difference between the charged and neutral $B$ decay suggested in \cite{Benzke:2010tq} as a probe of NP is not very promising in our scenario, since it probes the difference of the phases of $C_7$ and $C_8$, which are approximately aligned here.
}.

\subsubsection*{Angular CP asymmetries in $B\to K^*\mu^+\mu^-$}

The angular distribution of $B_d\to K^*(\to K\pi)\mu^+\mu^-$, which is measurable at the LHCb experiment, gives access to a large number of observables sensitive to new physics. Since the flavour of the initial $B$ meson can be inferred from the charges of the final-state hadrons, it is also straightforward to measure CP asymmetries. In the current framework, where new physics effects enter mainly via the magnetic and chromomagnetic operators, the most interesting effects arise in the T-odd CP asymmetries $A_7$ and $A_8$ \cite{Bobeth:2008ij,Altmannshofer:2008dz}.

Assuming new physics to enter only through the magnetic and chromomagnetic operators, while the chirality flipped and semileptonic operators are SM-like, the two normalized asymmetries, integrated in the low-$q^2$ range $1\div 6\;\text{GeV}^2$, can be written in the conventions of \cite{Altmannshofer:2008dz} as
\begin{align}
\langle A_7 \rangle \times R_\text{BR}
&\approx
-0.91 \, \text{Im}\!\left(C_7^\text{NP}\right)
 \,,
\label{eq:A7}
\\
\langle A_8 \rangle \times R_\text{BR}
&\approx
 0.51 \, \text{Im}\!\left(C_7^\text{NP}\right)
 \,,
\label{eq:A8}
\end{align}
where the tiny SM contribution has been neglected and the Wilson coefficients are to be evaluated at the scale $m_b$.
The factor $R_\text{BR}$ accounts for the modification of the CP-averaged branching ratio by the modification of $C_{7}$,

\begin{equation}
R_\text{BR} = \left.
\frac{\text{BR}(B_d\to K^*\mu^+\mu^-)+\text{BR}(\bar B_d\to \bar K^*\mu^+\mu^-)}{\text{BR}(B_d\to K^*\mu^+\mu^-)_\text{SM}+\text{BR}(\bar B_d\to \bar K^*\mu^+\mu^-)_\text{SM}}
\right|_{q^2\in[1,6]\,\text{GeV}^2}
\,.
\end{equation}
We note that, in constrast to the $b\to s\gamma$ case above, $R_{b\to s\gamma}$, this factor is not strongly constrained by experiment.

As a consequence of eqs. (\ref{eq:A7})--(\ref{eq:A8}), in models like the one at hand where NP enters only through $C_7$, there is a perfect correlation between $\langle A_7 \rangle$ and $\langle A_8 \rangle$ given by
\begin{equation}
\langle A_8 \rangle \approx -0.56 \;\langle A_7 \rangle \,.
\label{eq:A78}
\end{equation}

\subsubsection*{Time-dependent CP asymmetries in $B\to\phi K_S$ and $B\to\eta' K_S$}

The chromomagnetic operator also enters the hadronic penguin decays $B\to\phi K_S$ and $B\to\eta' K_S$.
The time-dependent CP asymmetries in these decays can be written as
\begin{equation}
\frac{\Gamma(B\to f)-\Gamma(\bar B\to f)}{\Gamma(B\to f)+\Gamma(\bar B\to f)}=S_f\sin(\Delta M t)-C_f\cos(\Delta M t)~.
\end{equation}
Denoting by $A_f$ ($\bar A_f$) the $B\to f$ ($\bar B\to f$) decay amplitude, $S_f$ and $C_f$ can be written as
\begin{equation}
S_f=\frac{2{\rm Im}(\lambda_f)}{1+|\lambda_f|^2} ~,~~
C_f=\frac{1-|\lambda_f|^2}{1+|\lambda_f|^2} ~,~~
\text{with}~~
\lambda_f=e^{-2i(\beta + \phi_{B_d})}(\bar{A}_f/A_f)~.
\end{equation}

In the SM, the mixing-induced CP asymmetries $S_{\phi K_S}$ and $S_{\eta' K_S}$ are predicted to be very close to $\sin2\beta$, measured from the tree-level decay $B\to J/\psi K_S$. In the presence of NP, there can either be a new contribution to the $B_d$ mixing phase, affecting both $S_{\phi,\eta' K_S}$ and $S_{\psi K_S}$, or a new loop contribution to the decay amplitude, affecting to a good approximation only $S_{\phi,\eta' K_S}$.

In the EMFV case, 
the new physics contribution $\phi_d$ to the $B_d$ mixing phase is very small and $S_f$ is modified only through the modification of the decay amplitude.
The dominant NP contribution to the decay amplitude comes from $C_8$. Then, it can be written as \cite{Buchalla:2005us,Hofer:2009xb}
\begin{equation}
A_f=A_f^c
\left[
1+a_f^ue^{i\gamma}+
\left(b_{f8}^c+b_{f8}^ue^{i\gamma}\right)C_8^{\rm NP *}
\right]\,,
\label{eq:Af}
\end{equation}
where the Wilson coefficient is to be evaluated at the scale $M_W$.
The $a_f$ and $b_f$ parameters can be found e.g. in \cite{Buchalla:2005us}.

\begin{figure}
\centering
\includegraphics[width=0.49\textwidth]{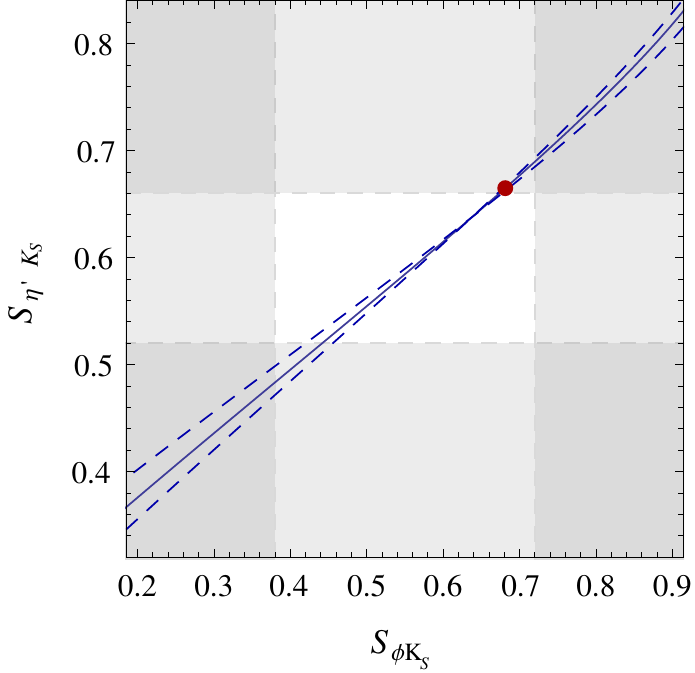}
\caption{Correlation between $S_{\phi K_S}$ and $S_{\eta' K_S}$ in models where NP enters only through $C_8$.
The solid line corresponds to $\text{Re}(C_8^\text{NP})=0$, the dashed lines to $\text{Re}(C_8^\text{NP})=\pm\text{Re}(C_8^\text{SM})$. The red point denotes the SM prediction. The gray areas correspond to the $1\sigma$ experimental bounds reported in table~\ref{tab:exp}.}
\label{fig:phieta}
\end{figure}

Since the deviations of $S_{\phi K_S}$ and $S_{\eta' K_S}$ from their SM values depend only on the Wilson coefficient $C_8$, there is a perfect correlation between them, shown in figure~\ref{fig:phieta} together with the $1\sigma$ experimental bounds.
The experimental data for both asymmetries have moved towards the SM value recently and are now compatible with SM at the $1\sigma$ level. Still, there clearly is room for NP, as shown in figure~\ref{fig:phieta}, and the asymmetries will be measured much more precisely in the future.

\section{Numerical analysis}
\label{sec:num}

For the numerical analysis, we vary the MSSM parameters at low energies, focusing on two benchmark cases:
\begin{enumerate}\renewcommand{\theenumi}{\roman{enumi}}
 \item An arbitrary phase for the $\mu$ term, but trilinear terms set to zero;
 \label{case:1}
 \item A real $\mu$ term, but an arbitrary phase in the (nonzero) stop trilinear coupling.
 \label{case:2}
\end{enumerate}
We stress again that case~\ref{case:2}.\ is equivalent to allowing arbitrary complex trilinears for all the sfermions, since the contributions to the observables decouple for all sfermions except the stop.
In both scenarios, we scan the MSSM parameters independently in the following ranges
\begin{align}
m_{\tilde q_3},m_{\tilde u_3},M_1,M_2,M_3,|\mu|,m_A &\in [200,700] ~\text{GeV,}\label{ml}
\\
m_{\tilde q_1},m_{\tilde q_2},m_{\tilde u_1},m_{\tilde q_2},m_{\tilde d},m_{\tilde \ell},m_{\tilde e} &\in [10,25] ~\text{TeV,}
\\
\tan\beta &\in [2,5] \,.
\end{align}
In case~\ref{case:1}., we scan $\phi_\mu$ from 0 to $2\pi$ and set $A_t=0$; in case~\ref{case:2}., we choose positive $\mu$ and
\begin{equation}
\frac{|A_t|}{m_{\tilde q_3}} \in [-3,3]
\,,\qquad
A_t = |A_t| e^{i\phi_{A_t}}
\,,\qquad
\phi_{A_t} \in [0,2\pi]
\,.
\end{equation}

We discard points violating sparticle mass bounds (in particular, the lightest stop mass is required to be above 95.7~GeV \cite{Nakamura:2010zzi}, which is relevant in scenario ~\ref{case:2}.) and are in disagreement with $\text{BR}(B\to X_s\gamma)$
or $\epsilon_K$ at more than $2\sigma$.
We calculate all the relevant quantities performing the full diagonalization of sparticle mass matrices, i.e. we are not making use of the mass insertion approximation employed in sections \ref{sec:edm} and \ref{sec:C78} to display the main dependence on SUSY parameters. We use a modified version of the \texttt{SUSY\_FLAVOR} code \cite{Rosiek:2010ug} to cross-check part of our results.

\begin{figure}
\begin{center}
\includegraphics[width=0.49\textwidth]{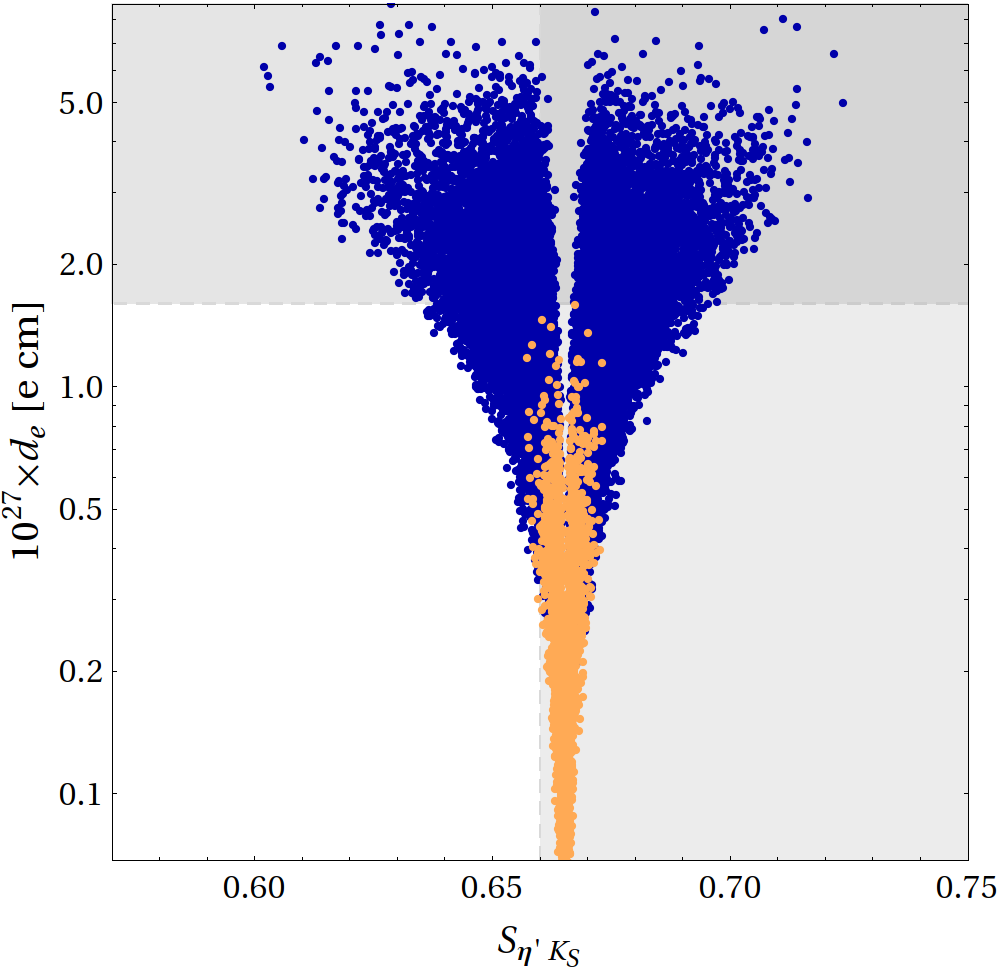}%
\hspace{0.019\textwidth}%
\includegraphics[width=0.49\textwidth]{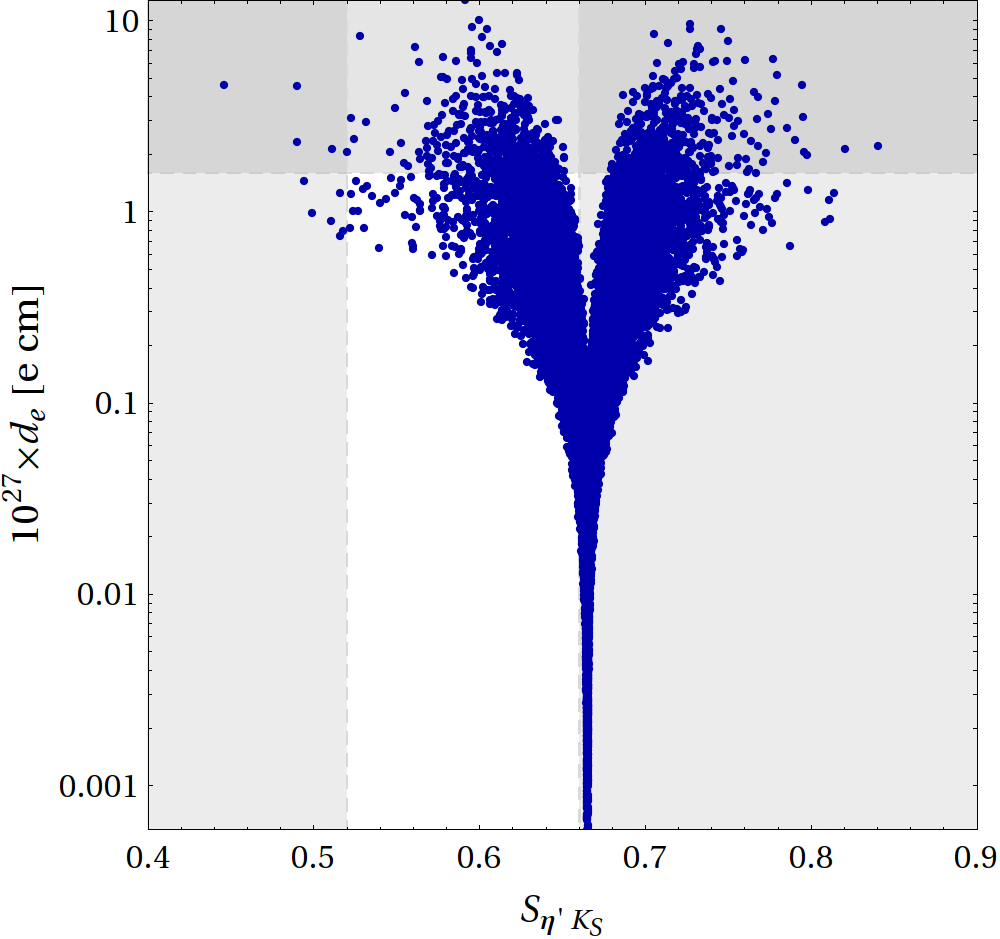}
\end{center}
\caption{Correlation between the electron EDM and the mixing-induced CP asymmetry in $B\to\eta' K_S$ in the two scenarios with a complex $\mu$ term (left) or complex $A_t$ term (right).
The gray areas indicate the 90\% C.L. upper bound in the case of $d_e$ and the experimental $1\sigma$ range in the case of $S_{\eta' K_S}$.
In the left-hand plot, the orange points have $|\sin\phi_\mu|<0.2$.
}
\label{fig:devsS}
\end{figure}

We now turn to the numerical analysis of the effects in EDMs and $B$ physics. In figure~\ref{fig:devsS}, we show the correlation between the electron EDM, arising mostly from the two-loop Barr-Zee contributions, and the mixing induced CP asymmetry in $B\to\eta' K_S$
in the two scenarios. The 90\% C.L. upper bound on $d_e$, cf. (\ref{eq:deexp}), as well as the $1\sigma$ experimental range for $S_{\eta'K_S}$ are shown as gray areas.
In scenario~\ref{case:1}., where $\mu$ is complex, $d_e$ constitutes a significant constraint on the parameter space. Note that  (\ref{ml}) scans the lower left corner of all the figures \ref{fig:edm-contours}.
As a consequence, $S_{\eta'K_S}$ deviates from its SM prediction by at most $\pm0.05$,
which might be visible at super flavour factories, but would require a better control of the SM theory uncertainties.
The orange points in the left-hand plot of figure~\ref{fig:devsS} show those points which have $|\sin\phi_\mu|<0.2$. This demonstrates that a mild condition on the size of $\phi_\mu$ is enough to always fulfill the EDM bounds; however, the resulting effects in $B$ physics are even smaller.
In scenario~\ref{case:2}., with a real $\mu$ term and complex trilinears, on the other hand, much larger effects in $S_{\eta'K_S}$ 
are compatible with $d_e$ such that the current data on $S_{\eta'K_S}$ already exclude part of the parameter space, even imposing them only at the $2\sigma$ level.

\begin{figure}
\begin{center}
\includegraphics[width=0.49\textwidth]{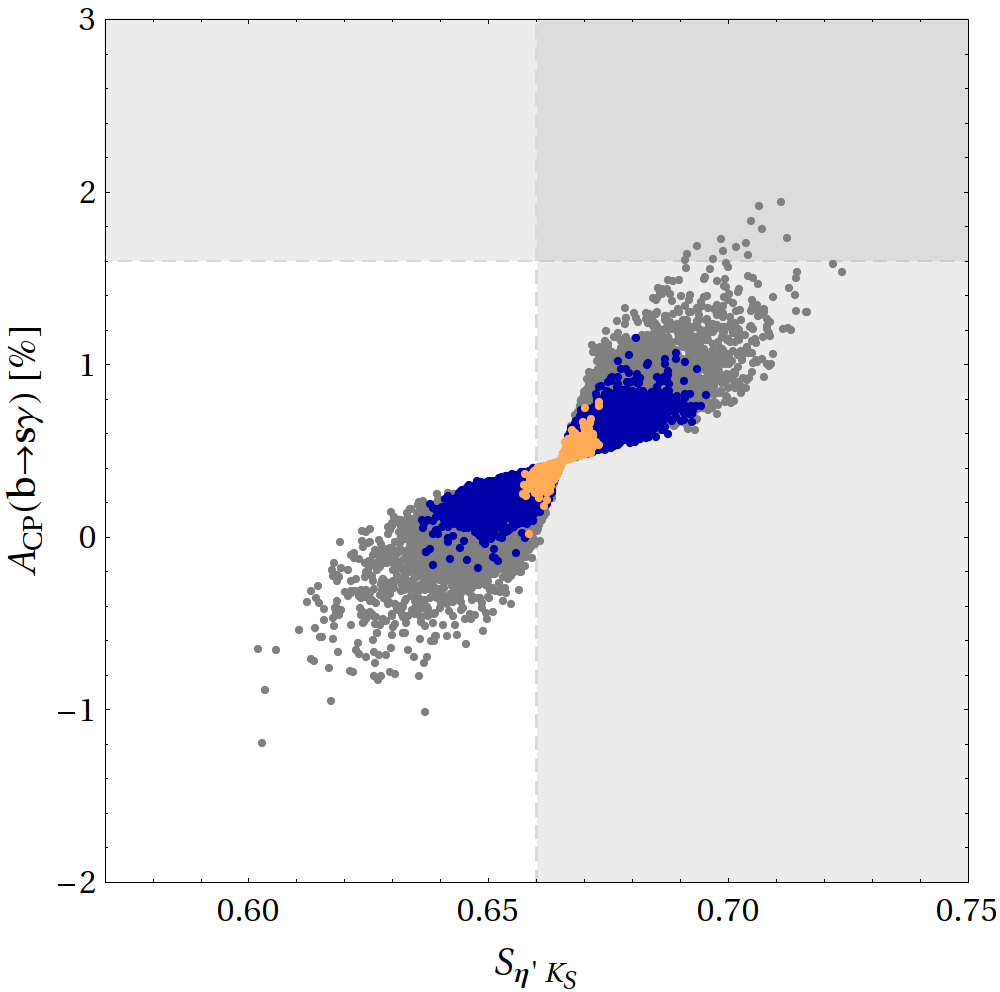}%
\hspace{0.019\textwidth}%
\includegraphics[width=0.49\textwidth]{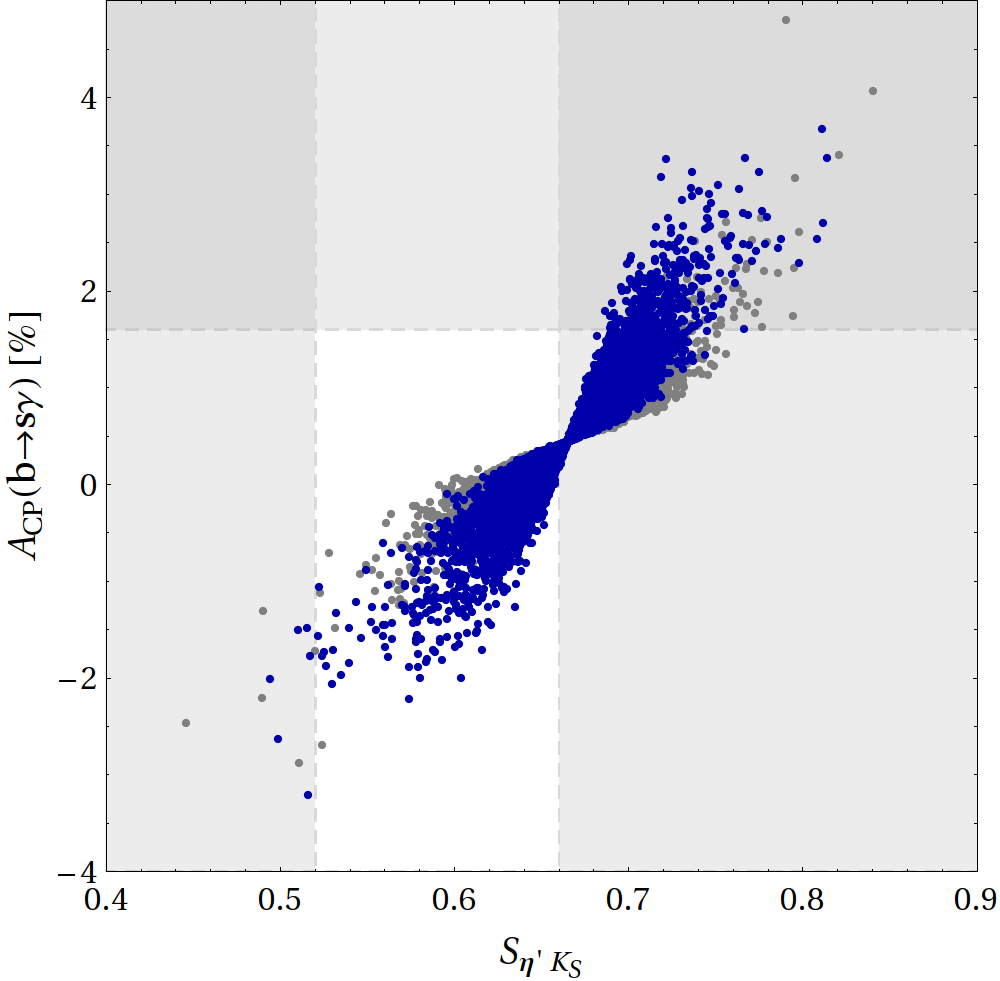}\\
\includegraphics[width=0.49\textwidth]{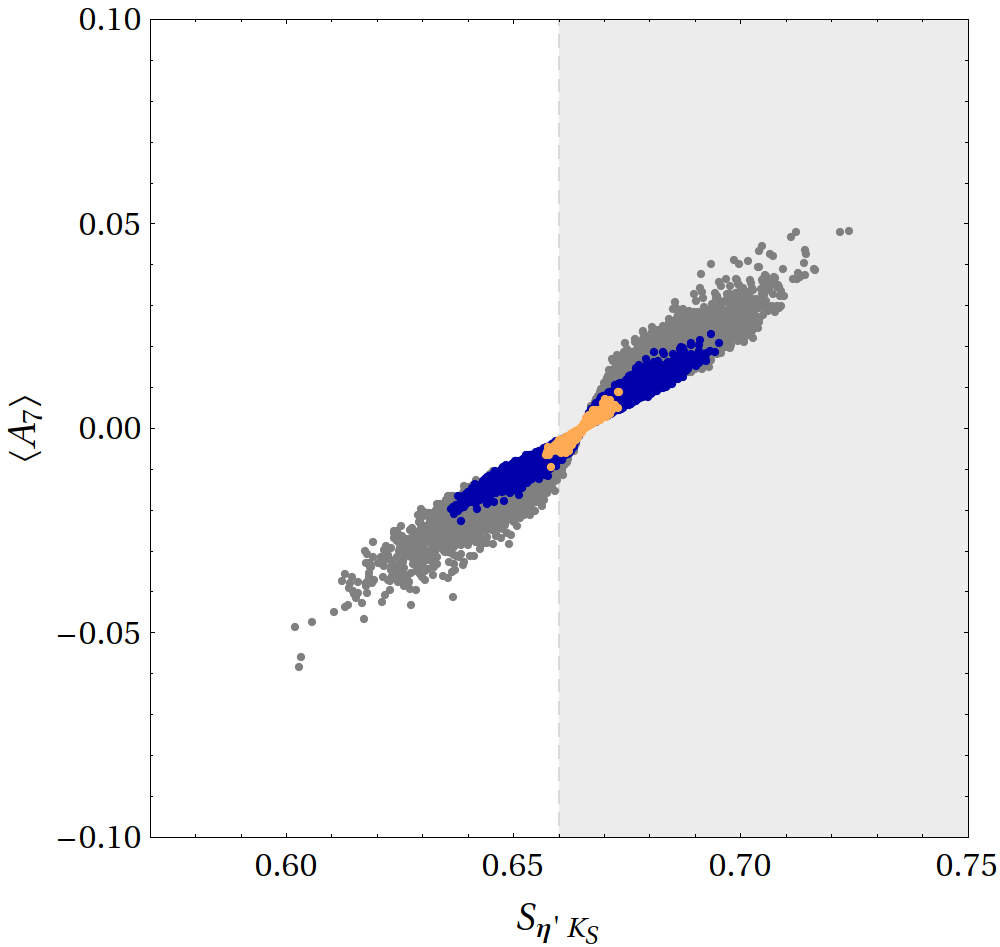}%
\hspace{0.019\textwidth}%
\includegraphics[width=0.49\textwidth]{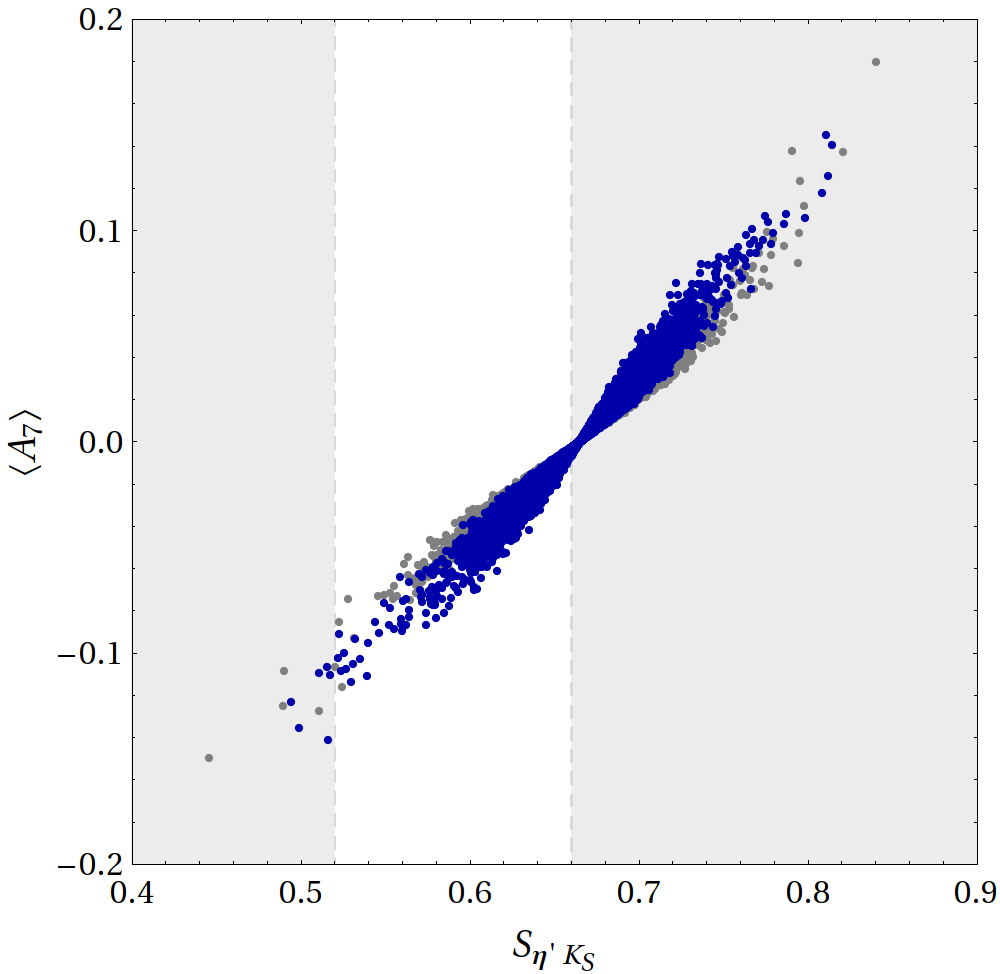}
\end{center}
\caption{Correlation between the mixing-induced CP asymmetry in  $B\to\eta' K_S$ and the direct CP asymmetry in $b\to s\gamma$ (top row) as well as the angular CP asymmetry $\langle A_7 \rangle$ in $B\to K^*\mu^+\mu^-$
in the two scenarios with a complex $\mu$ term (left column) or complex $A_t$ term (right column).
The gray points are allowed by all constraints except $d_e$, while the blue points are compatible with all constraints. The orange points in the left column have $|\sin\phi_\mu|<0.2$, as in fig.~\ref{fig:devsS}.}
\label{fig:SvsA}
\end{figure}

The same trend can be seen in figure~\ref{fig:SvsA}. In scenario~\ref{case:1}., the effects in the $b\to s\gamma$ direct CP asymmetry are rather limited and at the border of sensitivity of the super flavour factories. The effects in $B\to K^*\mu^+\mu^-$ are very likely beyond experimental reach. In scenario~\ref{case:2}., $A_\text{CP}(b\to s \gamma)$ can become sizable, 
such that the current constraint from $B$ factories becomes relevant.
The CP asymmetry $\langle A_7\rangle$ in $B\to K^*\mu^+\mu^-$ can reach up to $\pm15\%$.

We do not show the corresponding effects in $S_{\phi K_S}$ and $\langle A_8\rangle$, but they can be easily read off from the correlations in figure~\ref{fig:phieta} and eq.~(\ref{eq:A78}).

After showing the results for the two benchmark cases, let us comment on the more general case of phases present in both $\mu$ and $(\mu A_t)$. In that case, the correlations among the $B$ physics observables will be similar to the ones in scenario~\ref{case:2}.\ shown in the right-hand column of figure~\ref{fig:SvsA}, since we showed that sizable effects in $B$ physics are only obtained in the presence of complex trilinear couplings. The correlation between $d_e$ and the $B$ asymmetries, on the other hand, will change due to possible cancellations. For example, in the general case it is possible to have SM-like $S_{\eta'K_S}$ even for very large values of $d_e$.

Let us also comment on phases of gaugino masses, which have not been discussed so far. As discussed in section~\ref{sec:setup}, it is always possible, by appropriate field redefinitions, to choose a basis where the $b$ term and one of the gaugino masses is real. We choose this to be the Wino mass $M_2$. Concerning the gluino mass parameter $M_3$, in the decoupling limit of the heavy squarks, there is no one- or two-loop contribution to the EDMs involving the gluino, and the only gluino contribution to $C_7$ and $C_8$ not suppressed by the heavy masses is real, as shown in section~\ref{sec:C78}. Thus, the phase of $M_3$ is irrelevant. The phase of $M_1$, on the other hand, enters via neutralino contributions to the electron EDM and to $C_{7,8}$. However, the neutralino contributions are in general subleading with respect to the chargino ones and do not lead to qualitatively new effects. We thus conclude that the numerical results are valid even for the most general case of non-universal gaugino masses.

Finally, we wish to mention that the signals and correlations in flavour 
physics arising in scenario~\ref{case:2}.\ are very similiar to the effects in the MFV MSSM 
with a complex $A_t$ term and a real $\mu$ term 
\cite{Altmannshofer:2008hc,Altmannshofer:2009ne}. In the latter case however, 
one needs to assume real first generation $A$ terms, which can be spoiled by RG 
effects \cite{Paradisi:2009ey}. Of course, the two setups are easily 
distinguishable on the basis of their different spectrum.

\section{Summary and Conclusions}
\label{sec:concl}

If only the sfermions interacting with the Higgs system via the top Yukawa, i.e. the stops and the left-handed sbottom, are light while all the other sfermions have multi-TeV masses, the SUSY CP problem is ameliorated since the one-loop contributions to the electric dipole moments are suppressed.
Since a hierarchical sfermion spectrum is not enough to cure the SUSY flavour problem without additional assumptions on the flavour structure of soft terms, it is natural to combine hierarchical sfermions with the Minimal Flavour Violation principle, which provides a symmetry explanation of the good agreement of FCNC data with the SM, but does not address the SUSY CP problem.

In this work, we have analyzed CP violation in a SUSY model with hierarchical sfermions and an assumption on the breaking of the flavour symmetry in the squark sector leading to  EMFV in low-energy processes. We have shown that all the phases allowed by the flavour symmetry, in particular the phase of the $\mu$ term, the gaugino masses and the trilinear couplings, can be sizable without violating the EDM constraints.

We performed a numerical analysis of two benchmark scenarios, i.\ with a complex $\mu$ term and vanishing trilinear couplings and ii.\ with a real $\mu$ term and sizable complex trilinears. In both cases, two-loop contributions to the EDMs independent of the first two generation sfermion masses lead to an electron EDM which is in the ballpark of the current experimental upper bound. In addition, effects are generated in CP asymmetries in $B$ physics to be scrutinized by
forthcoming experiments, in particular the mixing-induced CP asymmetries in $B\to \phi K_S$ and $B\to \eta' K_S$, the direct CP asymmetry in $B\to X_s\gamma$ and the angular CP asymmetries $A_7$ and $A_8$ in $B\to K^*\mu^+\mu^-$. In scenario~i., these effects are 
quite limited. In scenario~ii.\ and in the general case of phases in $\mu$ and $\mu A_t$, the effects in $B$ physics are sizable and could lead to interesting signatures.

While the setup analyzed in this work by no means provides a theory of flavour or CP violation, we believe that it constitutes an example of a simple solution to the SUSY flavour and CP problems which is in accord with naturalness \cite{Barbieri:2010pd} and does lead to visible signatures in flavour physics. Thus, it reaffirms the necessity to search for electric dipole moments and CP violation in $B$ physics as complementary tools to the LHC.

\section*{Acknowledgements}

We thank Paride Paradisi, Wolfgang Altmannshofer and Oleg Lebedev for useful comments. This work was supported by the EU ITN ``Unification in the LHC Era'', contract PITN-GA-2009-237920 (UNILHC) and by 
MIUR under contract 2008XM9HLM.

\bibliographystyle{utphys}
\bibliography{cpv_emfv}

\end{document}